\documentclass[onecolumn,amsmath,amssymb,aps,prd,reprint, showpacs, notitlepage, nofootinbib,singlecolumn]{revtex4}
\usepackage[paperwidth=210mm,paperheight=297mm,centering,hmargin=2.5cm,vmargin=2.5cm]{geometry}
\usepackage{hyperref}
\usepackage{graphicx}
\usepackage{dcolumn}
\usepackage{url}
\usepackage{bm}
\usepackage{mathrsfs}
\usepackage{nicefrac}
\usepackage{comment}
\usepackage{tikz-cd}

\usepackage{xcolor}
\DeclareUnicodeCharacter{2212}{-}
\begin{document}
\title{Einstein's special relativity from a space-time duality principle}
\date{ }
\author{Vikramaditya Mondal}%
\email{vikramaditya.academics@gmail.com}
\address{School of Physical Sciences, Indian Association for the Cultivation of Science, Kolkata-700032, India}

\begin{abstract}
     Einstein based his special theory of relativity on two postulates: (a) physical laws appear the same in all inertial frames, and (b) the speed of light in vacuum is an observer-independent constant. However, it is already known that the principle of the constancy of the speed of light is redundant in constructing the special theory of relativity. Adhering to this idea, we show here that the form of the Lagrangian for a free relativistic particle can be derived if we consider a class of theories that remain invariant under a duality transformation between space and time intervals. Therefore, the special theory of relativity is interpreted as a theory that exhibits a duality symmetry with space intervals and time intervals being dual to each other. The Lorentz transformations are then simply deduced as the linear transformations between inertial observers that leave the (infinitesimal) action of the relativistic particles unchanged.
\end{abstract}
\textbf{\pacs{03.30.+p}}

\maketitle
\section{Introduction}
The special theory of relativity not only offers an incredibly accurate description of the world around us but also provides a fundamental viewpoint towards understanding our universe. For example, special relativity forms the foundation of Quantum Electrodynamics, which is considered to be the most accurate fundamental theory ever constructed \cite{kinoshita2014tenth,PhysRevD.97.036001}. In 1905, Albert Einstein, from outside the circle of academia, proposed to discard the aether theory in favor of his special theory of relativity \cite{einstein1905electrodynamics, einstein1905does} and consequently, changed the course of modern physics altogether. Since its origin more than one hundred years ago, the special theory of relativity has withstood and still withstanding rigorous experimental scrutiny \cite{zhang1997special,jacobson2002tev,mattingly2005modern,will2005special,jacobson2006lorentz,wolf2006recent,mattingly2008have}. Due to such incredible success, research on constructing the fundamental theory from the ground up with an alternative perspective or on pursuing a better understanding of the underlying mathematical structure still remains to be of significant interest. We show here that the mathematics describing free relativistic particles admits a particular duality symmetry---symmetry under the exchange of space and time intervals---and imposing such a symmetry in the Lagrangian formalism of the mechanics of the free particle implies that the underlying geometry of the space-time cannot be Euclidean; instead a Minkowskian geometry is necessary to be adopted.\par

Einstein begun the analysis in his first relativity paper \cite{einstein1905electrodynamics} with two postulates---the special principle of relativity and the principle of the constancy of the speed of light in vacuum---both of which he thought were necessary for drawing the conclusions of his theory. Formulating special relativity in this way, that is, basing the theory on a postulate regarding the speed of light, made it seem as though special relativity is only a mere consequence of the Maxwellian electrodynamics (this fact becomes evident in the title of the relativity paper itself, ``On the electrodynamics of moving bodies''). This has led to a series of research directed towards the question of whether it is possible to construct the special theory of relativity purely based on kinematical or geometrical arguments independently of material dynamics, \textit{i.e.,} the laws of electrodynamics, or simply without involving light. Minkowski's work revealed that special relativity combines space and time in a way that has applicability much beyond electrodynamics \cite{minkowski1909raum}. Even in the early works on special relativity, it was shown that the Lorentz transformations could be derived without the `constancy of the speed of light' postulate, that is, without any reference to the classical electrodynamics \cite{von1910einige,frank1911transformation,frank1912herleitung}. For more recent works that have argued the unnecessity of the second postulate to derive the Lorentz transformations, see the works \cite{schwartz1962axiomatic,lee1975lorentz, levy1976one,schwartz1984deduction,mermin1984relativity,schwartz1985simple,singh1986lorentz,sen1994galileo,Field_2001,pal2003nothing,coleman2003dual,field2004new, dunstan2008derivation,pelissetto2015getting,mathews2020seven} and the references therein (note that the list given here is not exhaustive).\par The derivation of Lorentz transformation inevitably involves considering several assumptions. For example, one, generally, starts with the assumption that space and time constitute a Euclidean manifold ($\mathbb{R}^3\times\mathbb{R}$) with vanishing intrinsic curvature. Such a manifold is supposed to have additional symmetries---the homogeneity of space and time (symmetry under four-translation) and the isotropy of space (symmetry under three-rotation). Moreover, one considers the special principle of relativity, \textit{i.e.}, the first postulate, which mathematically amounts to the statement that given two observers $S$ and $S^\prime$ where $S^\prime$ moves along their common axis with a velocity $v$ with respect to $S$, the linear transformation matrix $\Lambda$ which transforms the unprimed coordinates to the primed ones ($x^\prime=\Lambda x$) is only a function of the relative velocity, $\Lambda=\Lambda(v)$, and under the transformation $v\rightarrow-v$, the coordinate transformation matrix transforms to its inverse $\Lambda(-v)=\Lambda^{-1}$, that is to say, as the velocity of $S$ is $-v$ with respect to $S^\prime$ the coordinate transformation from primed to unprimed coordinate is obtained by replacing $-v$ in the places of $v$ in $\Lambda$, giving $x=\Lambda(-v)x^\prime$. The special principle of relativity, along with the symmetry of the manifold---homogeneity, and isotropy---is sufficient to land us into coordinate transformation laws which are consistent with both Galilean and Einsteinian relativity, thereby confirming their common origin from the principle of relativity (see for example in \cite{pal2003nothing,dunstan2008derivation,mathews2020seven}). Thereafter, in conventional approaches, the second postulate concerning the speed of light might be used to arrive at the Lorentz transformations finally. Whereas, in the alternative approaches, the second postulate is replaced by invoking some other physical or mathematical principles, like, Newton and Maxwell's laws in the case of \cite{dunstan2008derivation}; or group properties (such as closure) of the coordinate transformations in the case of \cite{schwartz1962axiomatic,lee1975lorentz,levy1976one} \textit{et cetera}. Given the Lorentz transformations, a notion of invariant four-dimensional distance can be constructed if we consider non-Euclidean geometry with Lorentzian signature. Therefore, the three-space ($\mathbb{R}^3$) and time ($\mathbb{R}$) can be combined into a four-dimensional Minkowskian ($\mathbb{M}^4$) manifold.\par
Our approach here is different. Instead of deriving the coordinate transformations directly, we derive the form of the Lagrangian for a free relativistic particle by imposing homogeneity, isotropy, and a principle of mathematical duality between space and time intervals, that is, the Lagrangian based description of the motion of the particle must remain invariant under the transformation,
\begin{equation}\label{eq:space-time_duality}
    \begin{gathered}
    |\Delta \mathbf{x}|\rightarrow -i~k~\Delta t,\\
    \Delta t\rightarrow i~k^{-1}~|\Delta \mathbf{x}|,
    \end{gathered}
\end{equation}
where, $|\Delta\mathbf{x}|$ is the spatial distance traveled by the particle in time interval $\Delta t$, and $i$ is the complex number $\sqrt{-1}$, making the transformation to be discrete and complex. Here, $k$ is a suitable finite constant of dimension $LT^{-1}$ which defines the scale for the duality.\par
This transformation is a testament to the fact that the Lorentzian line element, ${\rm d}s^2=-c^2{\rm d}t^2+|{\rm d}\mathbf{x}|^2$, remains invariant under the duality transformation mentioned above (with $k$ replaced with $c$). Our real goal is to show that an \textit{a priori} demand of the invariance of Lagrangian mechanics under this duality transformation \textit{uniquely} leads to special relativity. However, it must also be mentioned that the appearance of complex numbers in the transformation has nothing to do with the age old trick of defining the time coordinate as $x^4=ict$ in order to generate a sign difference between spatial and temporal part of the Euclidean metric (see, \cite{minkowski1909raum}). To show this concretely, if we start with a Euclidean metric, ${\rm d}s^2=k^2{\rm d}t^2+|{\rm d}\mathbf{x}|^2$, and then apply the duality transformation we get a negative definite Euclidean metric, ${\rm d}s^2=-k^2{\rm d}t^2-|{\rm d}\mathbf{x}|^2$, and \textit{not} the Minkowski metric. Finally, here,  There is no \textit{a priori} reason to consider that the scale of such a mathematical duality would be dependent on any special observer, and thus, $k$ should be a universal constant. This can be seen more concretely too. Consider two observers with their coordinates being primed and unprimed ones, and let them define the duality transformations with different constants primes and unprimed-$k$, then a series of alternating transformations ${\rm duality}\rightarrow {\rm coordinate} \rightarrow {\rm duality} \rightarrow {\rm coordinate}$, that is, \par
\begin{center}
   \begin{tikzcd}
    (\frac{k}{k^\prime} | {\rm d}\mathbf{x}|, \frac{k^\prime}{k}{\rm d}t ) & (|{\rm d} \mathbf{x}|,{\rm d}t) \arrow{rr}{{\rm duality}} & & (-i k {\rm d}t, i k^{-1} |{\rm d} \mathbf{x}|) \arrow{d}{{\rm coordinate}}\\
     & \arrow{ul}{{\rm coordinate}} (\frac{k}{k^\prime} | {\rm d}\mathbf{x^\prime}|, \frac{k^\prime}{k}{\rm d}t^\prime )  & & \arrow{ll}{{\rm duality}} (-i k {\rm d}t^\prime, i k^{-1} |{\rm d} \mathbf{x^\prime}|)
    \end{tikzcd} 
\end{center}
suggests that in order for the duality transformation to be consistent, we must have an universal constant $k$. This constant can be identified with the speed of light due to several theoretical or experimental facts, as we shall argue later. Now, as we are beginning with a universally constant velocity, it might seem that the principle of constancy of speed of light is somehow assumed from the beginning! However, so far, there is no connection is made for $k$ with the speed of light and neither it has any physical significance, \textit{e.g.}, it's not yet a universal speed limit on signal propagation. Having a universal constant of dimension of velocity does not automatically lands us onto a Minkwoskian geometry, that is to say, additionally, the symmetry of the particle's dynamics under the duality transformation must hold in order for this constant to have any significance at all. Another pertinent remark regarding the defined duality transformation is that duality transformations, in general, are always defined on dynamical degrees of freedom of the theory (for example, electromagnetic duality is defined on electric and magnetic fields). However, in our case we have demanded duality between quantities which are not directly the dynamical degrees of freedom in the theory of particle motion. This does not bother us for the reason that even though our dual quantities are not the dynamical degrees of freedom these are physically measurable quantities, that is, one can always physically measure distance and time intervals using rulers and clocks without even having to set up a coordinate system, which is an abstraction.\par
Physically, the free particles being free from any dynamical influence, trace out the geodesics of the underlying manifold. Therefore, the Lagrangian or the action describing such motion is quite sensitive to the geometry of the manifold, which, we shall see, will be revealed to be Minkowskian. Once we have obtained the form of the action, then the Lorentz transformations can be derived by demanding that the coordinate transformation between inertial frames must leave the action invariant by virtue of the principle of relativity as the action is a fundamental quantity responsible for the laws of motion of the particle.\par

In what follows, we shall briefly review, as examples, the duality symmetry in Maxwell's electromagnetic theory and thermodynamics in order to familiarize ourselves with the concept before proceeding to derive the form of the Lagrangian for a theory that admits duality between space and time intervals.

\section{Duality in electromagnetism and thermodynamics}

The concept of duality in physics is an old and important one. The origin of this idea goes back to the beginning days of electromagnetic theory \cite{heaviside2003electromagnetic,heaviside2011electrical,heaviside1892xi,larmorwork}. In recent times, duality principles have found significant applications in string theory and quantum gravity research (see, \textit{e.g.}, \cite{polchinski1996string, horowitz2009gauge, padmanabhan1997duality} and so on), making the idea worth exploring.\par
In different context duality may mean different things. Thus, for convenience, we define the sense in which we use the concept of a duality symmetry. Consider a theory is concerned with quantities $\{A,B,\dots\}$ and their dual quantities $\{X,Y,\dots\}$; then the theory is said to be duality symmetric if certain equations in the theory have mathematical structure such that those remain invariant under the duality transformations,
\begin{equation}
    \begin{gathered}
    A\rightarrow({\rm constant})({\rm scale})~X,~~X\rightarrow({\rm constant})({\rm scale})~A,\\
    B\rightarrow({\rm constant})({\rm scale})~Y,~~Y\rightarrow({\rm constant})({\rm scale})~B,\\
    \vdots
    \end{gathered}
\end{equation}
with appropriate constants (real or imaginary) and scales of the transformations. We shall now provide specific examples in the context of electromagnetic theory and thermodynamics.\par
Consider, the Maxwell equations in vacuum,
\begin{equation}
    \begin{gathered}
    \boldsymbol{\nabla}\boldsymbol{\cdot}\mathbf{E}=0,\\
    \boldsymbol{\nabla}\boldsymbol{\cdot}\mathbf{B}=0,\\
    \boldsymbol{\nabla}\times\mathbf{E}=-\frac{\partial\mathbf{B}}{\partial t},\\
    \boldsymbol{\nabla}\times\mathbf{B}=\mu_0\epsilon_0\frac{\partial\mathbf{E}}{\partial t}.
    \end{gathered}
\end{equation}
In this theory the physical quantities, the electric field $\mathbf{E}$ and the magnetic field $\mathbf{B}$ are dual with respect to each other. Then, it is quite easy to see that the Maxwell's equations, as a set, remain invariant under the dual transformations,
\begin{equation}
    \begin{gathered}
    \mathbf{E}\rightarrow-\frac{1}{\sqrt{\mu_0\epsilon_0}}~\mathbf{B},\\
    \mathbf{B}\rightarrow\sqrt{\mu_0\epsilon_0}~\mathbf{E}.
    \end{gathered}
\end{equation}
Here the constants are $\pm 1$, and the scale of the transformation is defined by the quantity $\frac{1}{\sqrt{\mu_0\epsilon_0}}$, which in this case corresponds to the speed of light in vacuum. Therefore, the vacuum Maxwell's equations exhibit a duality symmetry. Let us see another specific example in thermodynamics \cite{PhysRev.15.269}. Consider a reversible system at temperature $T$ characterized by $\delta Q$, the heat energy received by the system; $d S$, the corresponding change in the entropy; $d U$, the change in the internal energy; and $\delta W$, the work done by the system. Now as dual to $\delta Q$ define $\delta E$, which is the entropy measure of the energy given out by the system; and as dual to work done $\delta W$ define $\delta R$ as the entropy measure of the same work. Then we have the following set of thermodynamic relations,
\begin{equation}\label{eq:set_one}
    \begin{gathered}
    \delta Q=TdS,\\
    \delta Q-\delta W=dU,\\
    \delta R=\frac{\delta W}{T},\\
    \delta E=\frac{\delta W-\delta Q}{T}.
    \end{gathered}
\end{equation}
Let us now define the duality transformations between thermodynamic variables $\{Q,W,S,U,T\}$ to their duals $\{E,R,U,S,T\}$ as follows,
\begin{equation}\label{dual_transformation}
    \begin{gathered}
    Q\leftrightarrow E,~~ W\leftrightarrow R,~~ S\leftrightarrow-U,~~ T\leftrightarrow\frac{1}{T}.
    \end{gathered}
\end{equation}
With these transformations the set of equations (\ref{eq:set_one}) transforms to the following set of equations,
\begin{equation}
    \begin{gathered}
    \delta E=-\frac{dU}{T},\\
    \delta E-\delta R=-dS,\\
    \delta W=T\delta R,\\
    \delta Q=T\left(\delta R-\delta E\right),
    \end{gathered}
\end{equation}
which after rearrangement can be shown to be the same set of equations as (\ref{eq:set_one}) in changed order. Therefore, the thermodynamic equations (\ref{eq:set_one}) exhibits a duality symmetry under the dual transformations (\ref{dual_transformation}). The measures of energy are dual to the measures of entropy and temperature is self dual.\par

Similar to these examples, we shall show below that we can derive the Lagrangian for relativistic free particles by demanding that the theory is symmetric under the space-time duality transformation (\ref{eq:space-time_duality}) along with the homogeneity and isotropy symmetry. Then the Lorentz transformation follows as transformation between inertial frames, i.e., linear transformations that leave the infinitesimal action of the theory invariant.

\section{The Lagrangian for a space-time duality symmetric theory}

The evolution of a wide range of systems can be derived using the Lagrangian formulation. The textbooks generally remark that the Lagrangian is defined as kinetic energy minus the potential energy in particle mechanics. The reason why Lagrangians take that form is generally left undiscussed. Though the Lagrangian for a most general system might be arbitrary, it is possible to motivate the form of the Lagrangian by appealing to the symmetries the system is supposed to exhibit. For example, the form of the Lagrangian for a free non-relativistic particle can be motivated by appealing to the homogeneity and isotropy of the system and by demanding that the action remains invariant under Galilean transformation \cite{landau1976mechanics}. Similarly, the form of the Lagrangian for a relativistic free particle can be derived by demanding that the action for the system is invariant under Lorentz transformations. Here, we shall try to motivate the form of the Lagrangian for theories (of free particles) that remain invariant under the duality transformation given in (\ref{eq:space-time_duality}). As discussed earlier, such a Lagrangian or an action corresponding to free particles will be sensitive to the geometry of the underlying spacetime.\par
Let us consider a particle whose trajectory is described by a parametrized smooth curve in the three dimensional Euclidean space,
\begin{align}
    \mathbf{x}(t):\mathbb{R}\rightarrow\mathbb{R}^3.
\end{align}
Here, we have chosen the observer/coordinate time `$t$' (where, $t\in [a,b]$) as the parameter for the the trajectory. The curve, by assumption, is continuous and differentiable everywhere in the domain $[a,b]$. The motion of the particle is quantified by its instantaneous velocity at time $t\in[a,b]$, defined as the tangent vector to the curve at that point,
\begin{equation}\label{eq:velocity}
    \begin{gathered}
        \mathbf{v}(t):\mathbb{R}\rightarrow\mathbb{R}^3,\\
        \mathbf{v}(t)=\lim_{\Delta t\rightarrow 0}\frac{\mathbf{x}(t+\Delta t)-\mathbf{x}(t)}{\Delta t}.
    \end{gathered}
\end{equation}
In particle mechanics, it is postulated that the dynamics of the particle is such that the trajectory followed by the particle (the curve $\mathbf{x}(t)$ running between the instants $t=a$ to $t=b$) is the extremum point of a fundamental quantity called the action functional,
\begin{align}
    S[\mathbf{x}(t)]=\int_a^b L~{\rm d}t,
\end{align}
where, the Lagrangian $L$ is a function of the dynamical variable specifying the particle's position at an instant $\mathbf{x}(t)$, its velocity ${\mathbf{{v}}}(t)$ and maybe of time,
\begin{align}
    L=L\left(\mathbf{x},\mathbf{v},t\right).
\end{align}
The trajectory that extremizes the action satisfies Euler-Lagrange's equations,
\begin{align}
    \frac{\rm d}{{\rm d}t}\left(\frac{\partial L}{\partial \mathbf{v}}\right)=\frac{\partial L}{\partial \mathbf{x}},
\end{align}
which are the equations of motion satisfied by the particle. In non-relativistic mechanics, the Lagrangian is taken to be $L=T-V$, by convention, where $T$ is the particle's kinetic energy and $V$ is the potential energy (in the following, we shall be working with free particles only, and hence, the potential energy part of the Lagrangian will be ignored). In conventional textbooks, there is an aura of mystery regarding why should the Lagrangian must take such a form. However, it is well known that this form of the Lagrangian and also the form in the case of a relativistic particle can be derived using the fact that the action, which determines the dynamical laws must be invariant under the transformations between inertial frames, Galilean transformation in the case of non-relativistic particles and Lorentz transformation in the case of relativistic particles. In what follows, we shall determine the Lagrangian for the relativistic free particle by demanding symmetry under the duality transformation between space and time intervals along with some additional symmetry properties of the manifold on which the particle's trajectory is defined.\par 
For a free particle, the manifold on which its trajectory is defined has neither a special point in space nor a special instant in time; that is, space and time are homogeneous. The homogeneity of space and time then implies that the Lagrangian does not explicitly depend on the position $\mathbf{x}$ of the particle and time label `$t$'. Moreover, space has no special direction as well. The isotropy of space implies that the Lagrangian is independent of the direction of the particle's velocity $\mathbf{v}(t)$ and only depends on its magnitude, that is,
\begin{align}
    L=L\left(v^2\right).
\end{align}
Here, $v^2$ is the magnitude square of the velocity, defined as,
\begin{align}
    v^2=g\left(\mathbf{v},\mathbf{v}\right)=\mathbf{v}^Tg\mathbf{v},
\end{align}
with $g(~,~):\mathbb{R}^3\times \mathbb{R}^3\rightarrow\mathbb{R}$ being a symmetric, bi-linear, positive semi-definite metric on the Euclidean manifold $\mathbb{R}^3$, and for our purpose, we consider the Cartesian metric $g={\rm diag}(+1,+1,+1)$. The scalar product between any two vectors $\mathbf{A}$ and $\mathbf{B}$ is defined as
\begin{align}
    \mathbf{A}\boldsymbol{\cdot}\mathbf{B}=g(\mathbf{A},\mathbf{B})=\mathbf{A}^Tg\mathbf{B}.
\end{align}
As we have taken the Lagrangian to be independent of the position of the particle, the Euler-Lagrange equations reduce to,
\begin{align}\label{eq:euler-lagrange}
    \frac{\rm d}{{\rm d}t}\left(\frac{\partial L}{\partial \mathbf{v}}\right)=\mathbf{0}.
\end{align}\par
Now, we are interested in the theories which are invariant under the duality transformations between space and time intervals (\ref{eq:space-time_duality}), which essentially means that at any instat $t$, if the particle moves an infinitesimal arc-length ${\rm d}s$ along the curve in an infinitesimal time or parameter interval ${\rm d}t$, then the action and the Euler-Lagrange equations remain invariant under the following transformation of the tuple $({\rm d}s,k~{\rm d}t)\in\mathbb{R}^+\times\mathbb{R}^+$ into $(-i~k~{\rm d}t,i~{\rm d}s)\in i\mathbb{R}\times i\mathbb{R}$, that is,
\begin{align}\label{eq:duality_mathematical}
        \mathbb{R}^+\times\mathbb{R}^+\ni\begin{pmatrix}
            {\rm d}s\\
            k~{\rm d}t
        \end{pmatrix}
        \rightarrow &
        \begin{pmatrix}
            0 & -i\\
            i & 0
        \end{pmatrix}
        \begin{pmatrix}
            {\rm d}s\\
            k~{\rm d}t
        \end{pmatrix}\in i\mathbb{R}\times i\mathbb{R}\backslash\{\mathbf{0}\}\nonumber\\
        &\forall~ t\in \mathbb{R}.
\end{align}
First, we must understand how the action and the Euler-Lagrange equations change under such a duality transformation. To begin with, notice that the action is defined as a sum over infinitesimal quantities $L~{\rm d}t$ along the trajectory as the parametrization of the curve is done by the parameter $t$. After the duality transformation, then, the transformed action must be defined with the measure $i~k^{-1}~{\rm d}s$ instead of ${\rm d}t$, which means that before implementing the duality transformation, we must change the parametrization of the curve with respect to $s$ instead of $t$, or in other words, the curve now has to be parametrized by the arc length of the curve rather than the coordinate time, that is,
\begin{equation}
\begin{gathered}
    \mathbf{x}(s):\mathbb{R}\rightarrow \mathbb{R}^3,\\
    s=\int {\rm d}s=\int \sqrt{g({\rm d}\mathbf{x},{\rm d}\mathbf{x})}.
\end{gathered}
\end{equation}
Now, the duality transformation concerns with the exchange between the magnitude of the distance travelled by the particle and the time interval it takes to cover that distance. Thus, the transformation cannot directly be applied the definition of velocity vector (\ref{eq:velocity}) to obtain the duality transformed counterpart of velocity. However, if we consider the magnitude of velocity $v$, then we can implement the duality transformation to get its dual counterpart $v_{\rm dual}$,
\begin{align}
    v=\frac{\sqrt{g({\rm d}\mathbf{x},{\rm d}\mathbf{x})}}{{\rm d}t}\rightarrow v_{\rm dual}=\frac{-i~k~{\rm d}t}{i~k^{-1}\sqrt{g({\rm d}\mathbf{x},{\rm d}\mathbf{x})}}=-k^2v^{-1}.
\end{align}
To avoid any mathematical pathology, for the time being, we exclude the case $v=0$; that is, we only consider particles traveling with a non-vanishing velocity (magnitude) with respect to a given observer or a reference frame. It would be useful later for us to compute the following quantity,
\begin{align}\label{eq:chain-rule}
    \frac{{\rm d}v}{{\rm d}v_{\rm dual}}=k^{-2}v^2.
\end{align}
The Lagrangian is a function of the magnitude of velocity. If the magnitude of velocity transforms to a quantity $v_{\rm dual}$, then the duality transformed counterpart of the Lagrangian reads,
\begin{align}
    L(v^2)\rightarrow L(v_{\rm dual}^2)\equiv L_{\rm dual}.
\end{align}
Thus we see that the proposed duality transformation (\ref{eq:duality_mathematical}) is equivalent to reparametrizing the trajectory with respect to arc length $s$ and then at every parameter value ($t$ or $s(t)$) replacing the magnitude of the tangent to the curve with its scaled inverse ($v\rightarrow v_{\rm dual}=-k^2v^{-1}$) in the functional dependencies of the quantities like the Lagrangian. Then the duality transformed action reads,
\begin{align}
    S[\mathbf{x}(t)]=\int_a^b L~{\rm d}t\rightarrow S_{\rm dual}[\mathbf{x}(s)]=i\int_{s_i}^{s_f} L_{\rm dual}~k^{-1}{\rm d}s.
\end{align}
As the action is a fundamental quantity determining the motion of the particle, we demand that the action remains invariant under such a duality transformation, which is to say that the quantities $S$ and $S_{\rm dual}$ are equal to each other. We can write $S_{\rm dual}$ as,
\begin{align}
    S_{\rm dual}[\mathbf{x}(t)]=i\int_a^b L_{\rm dual}~k^{-1}\frac{{\rm d}s}{{\rm d}t}{\rm d}t.
\end{align}
Comparing this with the expression for $S[\mathbf{x}(t)]$, we have the following relation between the particle's Lagrangian and its dual counterpart
\begin{align}
    L_{\rm dual} = -i~k\frac{L}{\nicefrac{{\rm d}s}{{\rm d}t}}=-i~kL~v^{-1}.
\end{align}\par
Let us, now, turn our attention to the invariance of the Euler-Lagrange equations of motion under the duality transformations, a condition that will provide us with a differential equation involving the Lagrangian, allowing us to determine its form. We see that the Euler-Lagrange equations (\ref{eq:euler-lagrange}) deal with the velocity vector, and thus we cannot directly apply the duality transformation. We must find equivalent equations that contain the same information however deals with the magnitude of velocity. As the Lagrangian is a function of $v^2$, we should try to find an equation involving terms like $\frac{\partial L}{\partial v^2}$ instead of $\frac{\partial L}{\partial \mathbf{v}}$ as it is in (\ref{eq:euler-lagrange}). Then we consider the following scalar product,
\begin{align}
    g\left(\frac{\mathbf{v}}{v^2},\frac{{\rm d}}{{\rm d}t}\left(\frac{\partial {L}}{\partial\mathbf{v}}\right)\right)=0.
\end{align}
As the Lagrangian only depends on the magnitude of the velocity, the solution for the equation of motion is, $\mathbf{v}={\rm constant}$. If the equation of motion is satisfied then the above equation can be rewritten as,
\begin{align}\label{eq:euler_scalar}
    \frac{{\rm d}}{{\rm d}t}\left[g\left(\frac{\mathbf{v}}{v^2},\frac{\partial {L}}{\partial\mathbf{v}}\right)\right]=0.
\end{align}
Now, we know $L=L(v^2)$, then the partial derivative can be changed to a derivative with respect to $v^2$ using the chain rule,
\begin{align}
    \frac{\partial L(v^2)}{\partial\mathbf{v}}=\frac{\partial L}{\partial v^2}\frac{{\rm d}\left(\mathbf{v}\boldsymbol{\cdot}\mathbf{v}\right)}{{\rm d}\mathbf{v}}=2\mathbf{v}\frac{\partial L}{\partial v^2}.
\end{align}
Thus, we have, for the scalar product in (\ref{eq:euler_scalar}),
\begin{align}
    g\left(\frac{\mathbf{v}}{v^2},\frac{\partial {L}}{\partial\mathbf{v}}\right)=2\frac{\partial L}{\partial v^2}.
\end{align}
Therefore, the equation equivalent to the Euler-Lagrange equations useful for us is the following,
\begin{align}
    2\frac{{\rm d}}{{\rm d}t}\left(\frac{\partial {L}}{\partial{v}^2}\right)=0.
\end{align}
It is convenient to refer to the expression in the left hand side of the above equation as Euler-Lagrange function $F$. When we say that the theory of a free particle is invariant under the duality transformations, we mean, two quantities namely the action of the particle and the Euler-Lagrange function $F$ are invariant under such transformations. Now, implementing the duality transformation on $F$, we get,
\begin{align}\label{eq:duality_acc}
    2\frac{\rm d}{{\rm d}t}\left(\frac{\partial L}{\partial v^2}\right)\rightarrow -2i~k\frac{{\rm d}}{{\rm d}s}\left(\frac{\partial L_{\rm dual}}{\partial v^2_{\rm dual}}\right),
\end{align}
which are both numerically equal to zero. We can simplify the expression in the right hand side as follows,
\begin{align}
    -2i~k\frac{{\rm d}}{{\rm d}s}\left(\frac{\partial L_{\rm dual}}{\partial v^2_{\rm dual}}\right)&=-2i~k\frac{{\rm d}}{{\rm d}s}\left(\frac{1}{2}v_{\rm dual}^{-1}\frac{\partial L_{\rm dual}}{\partial v_{\rm dual}}\right)\nonumber\\
    &=-i~k~v_{\rm dual}^{-1}\frac{\rm d}{{\rm d}s}\left(\frac{\partial L_{\rm dual}}{\partial v_{\rm dual}}\right)
\end{align}
Here, we have used the fact that $v$ or equivalently $v_{\rm dual}$ is a constant as long as the equations of motion are satisfied to pull the $v_{\rm dual}^{-1}$ outside the $s$ derivative. We can further simplify the above relation as
\begin{align}
    -2i~k\frac{{\rm d}}{{\rm d}s}\left(\frac{\partial L_{\rm dual}}{\partial v^2_{\rm dual}}\right)&=i~k^{-1}\frac{{\rm d}s}{{\rm d}t}\frac{\rm d}{{\rm d}s}\left(\frac{\partial L_{\rm dual}}{\partial v_{\rm dual}}\right)\nonumber\\
    &=i~k^{-1}\frac{\rm d}{{\rm d}t}\left(\frac{\partial \left(-i~kL~v^{-1}\right)}{\partial v}\frac{{\rm d}v}{{\rm d}v_{\rm dual}}\right)\nonumber\\
    &=k^{-2}\frac{\rm d}{{\rm d}t}\left(v^2\frac{\partial \left(L~v^{-1}\right)}{\partial v}\right),
\end{align}
here, in the last line we have used the relation (\ref{eq:chain-rule}). Now, the quantity inside the round parenthesis can be simplified further,
\begin{align}
    v^2\frac{\partial \left(L~v^{-1}\right)}{\partial v}&=v^2\left(-v^{-2}L+v^{-1}\frac{\partial L}{\partial v}\right)\nonumber\\
    &=2v^2\frac{\partial L}{\partial v^2}-L.
\end{align}
Therefore, with this manipulation the duality transformation of the Euler-Lagrange function $F$ (\ref{eq:duality_acc}) reads,
\begin{align}
    2\frac{\rm d}{{\rm d}t}\left(\frac{\partial L}{\partial v^2}\right)\rightarrow k^{-2}\frac{{\rm d}}{{\rm d}t}\left(2v^2\frac{\partial L}{\partial v^2}-L\right).
\end{align}
Then the duality symmetry implies that the above two expressions (which are each numerically equal to zero) are equal to each other, that is,
\begin{align}
    2\frac{\rm d}{{\rm d}t}\left(\frac{\partial L}{\partial v^2}\right)=k^{-2}\frac{{\rm d}}{{\rm d}t}\left(2v^2\frac{\partial L}{\partial v^2}-L\right).
\end{align}
Now, by rearranging the terms to one side we get,
\begin{align}
    \frac{{\rm d}}{{\rm d}t}\left[2\left(1-k^{-2}{v}^2\right)\frac{\partial L}{\partial{v}^2}+k^{-2}L\right]=0.
\end{align}
Thus, theories that are invariant under the duality transformation between space and time intervals (\ref{eq:duality_mathematical}) have Lagrangians of the form such that the following differential equation is satisfied,
\begin{align}
    \frac{d L}{d z}+\frac{k^{-2}}{2\left(1-k^{-2}z\right)}L+\frac{C}{2\left(1-k^{-2}z\right)}=0,
\end{align}
where we have set $v^2=z$ and $C$ is a constant, which, as we shall see, plays no significant role in the form of the action. The above equation is a first order inhomogeneous linear differential equation and can be solved by the method of integrating factor. The integration factor for the above differential equation is,
\begin{align}
    \exp\left(\int\frac{k^{-2}}{2(1-k^{-2}z)}dz\right)&=\exp\left(-\frac{1}{2}\log\left|1-k^{-2}z\right|\right)\nonumber\\
    &=\frac{1}{\sqrt{1-k^{-2}z}}.
\end{align}
Multiplying both sides of the differential equation with the integration factor we get,
\begin{align}
    \frac{{\rm d}}{{\rm d}z}\left[\frac{L}{\sqrt{1-k^{-2}z}}\right]=-\frac{C}{2\left(1-k^{-2}z\right)^{\frac{3}{2}}}.
\end{align}
Integrating both sides with respect to $z$, we finally get the solution for $L$ to be,
\begin{align}
    L(z)=Ck^2+D\sqrt{1-k^{-2}z},
\end{align}
where $D$ is an integration constant. As a constant term in the Lagrangian does not contribute in the equation of motion we can safely ignore it. Therefore, the Lagrangain has the following form,
\begin{align}
    L(v^2)=D\sqrt{1-k^{-2}v^2}.
\end{align}
We see that this above Lagrangian is radically different from the ordinary Lagrangian in the case of Newtonian mechanics, that is $\frac{1}{2}mv^2$. It is noticeable that if we perform a binomial expansion of the above Lagrangian in the limit $\frac{v}{k}\rightarrow 0$, then the leading order term with velocity (magnitude) dependence is of the order $v^2$,
\begin{align}\label{eq:lagrangian_expansion}
    L(v^2)\approx D-\frac{D}{2k^2}v^2+\cdots .
\end{align}
Therefore, we see that the Newtonian theory corresponds to the $\frac{v}{k}\rightarrow 0$ limit of the new duality symmetric theory. As the leading order constant term in the Lagrangian in (\ref{eq:lagrangian_expansion}) can be ignored, the requirement that the duality symmetric Lagrangian reduces to $\frac{1}{2}mv^2$ in this limit sets the constant $D$ to be the following,
\begin{align}
    D=-mk^2.
\end{align}
Therefore, finally, the Lagrangian of a massive free particle exhibiting duality symmetry has the form,
\begin{align}\label{eq:Lagrangian}
    L(v^2)=-mk^2\sqrt{1-\frac{v^2}{k^2}}.
\end{align}
We note that a set of different transformations, in which the signs are flipped with respect to (\ref{eq:duality_mathematical}), that is,
\begin{equation}\label{eq:duality_alternative}
    \begin{gathered}
        \begin{pmatrix}
            {\rm d}s\\
            k~{\rm d}t
        \end{pmatrix}\rightarrow
        \begin{pmatrix}
            0 & i\\
            -i & 0
        \end{pmatrix}
        \begin{pmatrix}
            {\rm d}s\\
            k~{\rm d}t
        \end{pmatrix},
    \end{gathered}
\end{equation}
also reproduces the same Lagrangian. Thus, only the relative sign of ${\rm d}s$ and $k~{\rm d}t$ after the duality transformations matters.\par The conjugate three momenta corresponding to the Lagrangian is
\begin{align}
    \mathbf{p}=\frac{\partial L}{\partial\mathbf{v}}=\frac{m\mathbf{v}}{\sqrt{1-\frac{v^2}{k^2}}}.
\end{align}
Then the Hamiltonian \footnote{This is only the zeroth component of the momentum four-vector. The covariant Hamiltonian of a relativistic particle vanishes due to its reparametrization invariance.} or energy of the particle is given by,
\begin{align}\label{eq:energy}
    H&=g\left(\mathbf{p},\mathbf{v}\right)-L\nonumber\\
    &=\frac{mk^2}{\sqrt{1-\frac{v^2}{k^2}}}.
\end{align}
We see that all the relevant expressions are consistent even when $v=0$ and thus our initial assumption of $v>0$ has no bearing on the final results. Now, it is clear from (\ref{eq:energy}) that, unlike in the Newtonian mechanics, the energy of the particle in its rest frame does not vanish; for $v=0$, we have,
\begin{align}
    \mathcal{E}=mk^2.
\end{align}
Subtracting out this rest energy from the total energy of the particle represents the energy of the particle only due to its motion, \textit{i.e.,} the particles kinetic energy,
\begin{align}\label{eq:kineticenergy}
    K=H-\mathcal{E}=mk^2\left[\left(1-\frac{v^2}{k^2}\right)^{-\frac{1}{2}}-1\right].
\end{align}
From the above expression, it is a well-known argument \cite{resnick1971introduction} that to accelerate a particle up to the finite speed $k$, it would require an infinite amount of energy to be provided to the particle; the impossibility of such a situation establishes $k$ as the universal speed limit in the universe. If such a limit did not exist, \textit{i.e.,} if $k\rightarrow\infty$, then we get back the Newtonian theory and $K$ becomes $\frac{1}{2}mv^2$ as expected.
\section{Lorentz transformations}
From the Lagrangian (\ref{eq:Lagrangian}) we can figure out the duality-symmetric infinitesimal action, which then reads,
\begin{align}\label{eq:infinitesimal_action}
    {\rm d}S=L(v^2){\rm d}t=-mk\sqrt{k^2{\rm d}t^2-|{\rm d}\mathbf{x}|^2}.
\end{align}
It is easy to see that the above action is indeed invariant under the transformations (\ref{eq:duality_mathematical}) or (\ref{eq:duality_alternative}). So far, we have not talked about coordinate transformation between inertial frames. All the arguments provided, so far, concerns only a single observer. The form of the Lagrangian is completely determined from the symmetry properties of the manifold, homogeneity and isotropy, and an extra principle of duality symmetry between space and time intervals measured by the said observer. Now, we can use the first postulate, which says the laws of physics appear the same in all inertial frames, to determine the valid coordinate transformation laws between the inertial frames. The infinitesimal action (\ref{eq:infinitesimal_action}) is a fundamental quantity determining the dynamics of the particle; thus, we demand that this quantity should remain invariant under the transformation between inertial frames. Demanding the invariance of the (infinitesimal) action is equivalent to demanding the invariance of the following quantity (as $m$ and $k$ are constants),
\begin{align}\label{eq:minkowski}
    -k^2{\rm d}t^2+|{\rm d}\mathbf{x}|^2\equiv \eta\left({\rm d}x,{\rm d}x\right),
\end{align}
where, the components of ${\rm d}x$ are ${\rm d}x^\mu=\left(k~{\rm d}t,{\rm d}\mathbf{x}\right)$ and $\eta(~,~)$ is the Minkowskian metric with mostly positive signature, $\eta={\rm diag}(-1,+1,+1,+1)$. Now, suppose there are two inertial observers $S$ and $S^\prime$, where $S^\prime$ is moving with a constant velocity $\mathbf{V}$ with respect to $S$, and there is a linear transformation \cite{eisenberg1967necessity} between their coordinates, like below,
\begin{align}
    x\rightarrow x^\prime=\Lambda x.
\end{align}
The quantity $\eta\left({\rm d}x,{\rm d}x\right)$ changes under the coordiante transformations to,
\begin{align}
    \eta\left({\rm d}x,{\rm d}x\right)\rightarrow\eta\left({\rm d}x^\prime,{\rm d}x^\prime\right)&=\eta\left(\Lambda{\rm d}x,\Lambda{\rm d}x\right)={\rm d}x^T\Lambda^T\eta\Lambda{\rm d}x\nonumber\\
    &=\Lambda^T\eta\Lambda\left({\rm d}x,{\rm d}x\right)
\end{align}
Then, the invariance of the infinitesimal action, i.e.,
\begin{align}
    \eta\left({\rm d}x^\prime,{\rm d}x^\prime\right)=\eta\left({\rm d}x,{\rm d}x\right),
\end{align}
implies, the following relation for the coordinate transformations is true,
\begin{align}\label{eq:condition}
    \Lambda^T\eta\Lambda=\eta.
\end{align}
It is well known that the Lorentz transformations,
\begin{equation}\label{eq:LT}
    \begin{gathered}
        \Lambda^0_{~0}=\gamma,~\Lambda^i_{~0}=-\gamma \frac{V^i}{k},~\Lambda^0_{~j}=-\gamma \frac{V_j}{k},\\
        \Lambda^i_{~j}=\delta^i_{~j}+V^iV_j\frac{\gamma-1}{V^2};\\
        {\rm where,}~\gamma=\frac{1}{\sqrt{1-\frac{V^2}{k^2}}},
    \end{gathered}
\end{equation}
are a possible solution to the equations (\ref{eq:condition}), see in \cite{weinberg1972gravitation}. The obvious deviation of these formulas from the conventional case is that instead of $c$, we have a finite constant $k$ which, so far, has no apparent connection to the speed of light as we have derived these transformations without the second postulate or any reference to the classical electrodynamics. The value of the free parameter $k$ that parametrizes a class of space-time duality symmetric theories can be set in several ways. For example, we can set up an experiment to measure the value of $k$ directly similar to the one suggested in section VII of \cite{mathews2020seven}. However, that is unnecessary as we can deduce the value of $k$ from the consistency requirement. The special relativity being fundamental is thoroughly absorbed into the framework of modern physics. All the results from high precision measurements---such as measurements in nuclear and particle physics, spectroscopy, measurements involving atomic clocks, say in the clock-comparison experiments or the GPS tracking system---can be explained harmoniously if the model with $k=c$ is adopted. Moreover, we might be able to fit formulas like (\ref{eq:kineticenergy}) against the data concerning radioactive decay, thereby indirectly measure the value of $k$ to be $c$ with appropriately designed experiments. Though not necessarily but in principle, if we were to refer to the classical electrodynamics and demand that the Maxwell equations retain their form under the transformations (\ref{eq:LT}) by virtue of the `principle of relativity' that the physical laws appear to be the same in all inertial frames, then $k$ can be readily identified with $c$.

\section{Relation to prior works}
The connection of the Lorentz symmetry group with the symmetry under space-time exchange has been investigated before as well. For example, in somewhat of a different context, in lattice quantum field theory, one starts with a four-dimensional Euclidean space(time) lattice which satisfies hypercubic point symmetry, that is, the symmetry of the lattice under the signed permutations of its coordinates $(\tau,x,y,z)$. Simply put, the transformation of the lattice points onto themselves with any of the coordinates, including the Euclidean time coordinate, being exchanged among themselves (with appropriate sign), preserves the physics of the system. The elements of such a group can be generated from parity transformations, and a discrete rotation with angle $\frac{\pi}{2}$ around the coordinate axes \cite{hamermesh2012group}. A scalar field on such a lattice should have a kinetic term like $M^{\mu\nu}\partial_\mu\phi\partial_\nu\phi$, where the matrix must have the hypercubic symmetry. In four dimensions, such a matrix is a constant multiple of $\delta^{\mu\nu}$. Hence, in the infrared or continuum limit, the hypercubic symmetry corresponds to the symmetry under the Euclidean $O(4)$ group. Thereafter, if we go back to real time from the Euclidean one via Wick rotation, i.e., with the transformation $\tau\rightarrow it$, then, with an additional condition of spacetime volume preservation, the symmetry group becomes $SO(3,1)$, implying Lorentz invariance (for more discussion, see \cite{mattingly2005modern,moore2003informal}).\par 
Our approach of space-time duality, presented here, is different for at least two main reasons---(a) we do not consider the transformation between individual coordinates $(t,x,y,z)$; instead, we work with a complex transformation between the infinitesimal distance (which, unlike the former case, requires additional metric information to convert coordinate separations into a physical distance $\sqrt{g({\rm d}\mathbf{x},{\rm d}\mathbf{x})}$) traveled by the particle in an infinitesimal time interval. Apart from being mathematically different, our transformation is even physically different from the hypercubic symmetry of the lattice, in the sense that the transformation we propose is concerned with the trajectory of the test particles (or geodesics) and not the coordinate system itself. Later, the information of the underlying geometry is extracted from the sensitivity of the test particle's trajectory to this geometry. This essentially means that the hypercubic symmetry between space and time is kinematical (\textit{i.e.}, geometrical) in nature (concerned with the (signed) permutation of the coordinates). In contrast, our duality symmetry between space-time is dynamical (concerned with the exchange of differentials $({\rm d}s,{\rm d}t)$, quantifying infinitesimal motion of the particles at all instants of the trajectory's parameter), making the two approaches fundamentally different from each other. (b) The consideration from hypercubic symmetry only lands us on the $O(4)$ group. To get to the $SO(3,1)$ group, we must apply an \textit{ad hoc} Wick rotation in addition to the space-time exchange symmetry (which is also equivalent to using the knowledge that spacetime metric has a Lorentzian signature). Nevertheless, in our case, the duality transformation is such that it uniquely determines the spacetime geometry to be Minkowskian. Therefore, though the hypercubic symmetry argument provides us with a wonderful insight regarding why Lorentz symmetry is an `accidental symmetry' of nature in the low energies while such symmetries might not be there in high energies, our argument here has advantages as far as deriving special relativity or Lorentz symmetry from least amount of assumptions (for example, we do not need to assume the existence of a fundamental scalar field) is concerned.\par
Another pertinent work \cite{Field_2001} proposes to derive the Lorentz transformations in a similar vein by positing a symmetry principle for exchange between space and time coordinates. However, as mentioned earlier, our approach with the duality transformation is quite different both mathematically and physically as we do not work with the transformation between individual spatial coordinates with the time coordinate; instead we work with transformations between the quantifiers of dynamics rather than that of kinematics. The derivation presented here, as far as we know, is original.

\section{Discussion}
Consider that an observer measures the infinitesimal distance ${\rm d}s=\sqrt{g\left({\rm d}\mathbf{x},{\rm d}\mathbf{x}\right)}$ travelled by a free particle in an infinitesimal time interval ${\rm d}t$ using the array of `rods' and `clocks', respectively, it has set up to form its coordinate system. Then, in the Lagrangian mechanical description of the particle's motion, if the observer switches the measurements made by the `rods' (${\rm d}s$) to that with the `clocks' (${\rm d}t$) associated with some complex factors, as in (\ref{eq:duality_mathematical}) or (\ref{eq:duality_alternative}), the invariance of the description of motion uniquely determines the form of the action. As we are considering free particles, their motion traces out the geodesics of the underlying manifolds. Therefore, the action of free particles contains the information of the kinematical or geometrical structure of the spacetime through which the particles traverse. We see that the infinitesimal action allows us to combine space and time into a four-dimensional manifold with Lorentzian signature as evident in (\ref{eq:minkowski}). Then the Lorentz transformations are simply the linear transformations between coordinates that preserve the action. Thus, the special theory of relativity is constructed here from a symmetry principle regarding the underlying mathematical structure of the theory. For the sake of clarity, we list the assumptions that were necessary for our derivation---(a) the homogeneity of space and time, (b) the isotropy of space, (c) the space-time duality symmetry in the Lagrangian based description of the particle's motion, and (d) the special principle of relativity. Unlike many other alternative derivations of Lorentz transformations, we did not have to use the mathematical version of the principle of relativity, that is ``$v\rightarrow-v$ implies $\Lambda(v)\rightarrow\Lambda(-v)=\Lambda^{-1}(v)$'', and, additionally, the fact that we are working with Lagrangian formulation (\textit{i.e.}, dynamics), sets our derivation apart from most of the works (with arguments from kinematics) done on the issue of deriving Lorentz transformations avoiding the second postulate.
Moreover, as we have not used the second postulate of special relativity, instead of the speed of light $c$, we have a universal duality scale $k$, which acts as the maximum speed limit in the universe and can be fixed with appropriate experiments. The arguments presented here pertaining to the duality symmetry only extend to free particles, just like the duality symmetry breaks down for Maxwell's equations with arbitrary charge and current distribution (as magnetic monopoles do not exist). However, by the analogy of the free particle case, the form of the Lagrangian for a relativistic particle moving in a potential $V(\mathbf{x})$ can be easily guessed. As in the $\frac{v}{k}\rightarrow 0$ limit, the Lagrangian of the free particle becomes of the order of kinetic energy (ignoring the constant term), the potential energy of the particle should be added (or subtracted) to the Lagrangian in the case where the particle is influenced by a potential. It is generally subtracted to match the convention of defining force as the `negative' gradient of the potential. Moreover, if the particle moves in a potential, the constant velocity $v$ should be replaced by $v(t)$ in the Lagrangian, giving us the usual form of the Lagrangian of a relativistic particle in a potential. This task of replacing $v$ with $v(t)$ can be easily done by means of a similar process of using a set of momentarily comoving reference frames as one usually does in the case of considering accelerating test particles in the framework of special relativity.
\section*{Acknowledgement}
Research of V.M. is funded by the INSPIRE fellowship from the DST, Government of India (Reg. No. DST/INSPIRE/03/2019/001887). V.M. would like to thank Dipayan Mukherjee for carefully reading the manuscript and providing several indispensable suggestions which have enriched the presentation of the article.




\begin{thebibliography}{}

\bibitem{kinoshita2014tenth}
Toichiro Kinoshita.
\newblock Tenth-order {QED} contribution to the electron g- 2 and high precision test of quantum electrodynamics.
\newblock In {\em Proceedings of the conference in Honour of the 90th Birthday
  of Freeman Dyson}, pages 148--172. World Scientific, 2014.
\bibitem{PhysRevD.97.036001}
Tatsumi Aoyama, Toichiro Kinoshita, and Makiko Nio.
\newblock Revised and improved value of the {QED} tenth-order electron anomalous
  magnetic moment.
\newblock {\em Phys. Rev. D}, 97:036001, Feb 2018.
\bibitem{einstein1905electrodynamics}
Albert Einstein.
\newblock On the electrodynamics of moving bodies.
\newblock {\em Annalen der physik}, 17(10):891--921, 1905.
\bibitem{einstein1905does}
Albert Einstein.
\newblock Does the inertia of a body depend upon its energy-content?
\newblock {\em Annalen der physik}, 18:639--641, 1905.
\bibitem{zhang1997special}
Yuan-Zhong Zhang.
\newblock {\em Special relativity and its experimental foundation}, volume~4.
\newblock World Scientific, 1997.
\bibitem{jacobson2002tev}
Ted Jacobson, Stefano Liberati, and David Mattingly.
\newblock Tev astrophysics constraints on Planck scale Lorentz violation.
\newblock {\em Physical Review D}, 66(8):081302, 2002.
\bibitem{mattingly2005modern}
David Mattingly.
\newblock Modern tests of lorentz invariance.
\newblock {\em Living Reviews in relativity}, 8(1):5, 2005.
\bibitem{will2005special}
Clifford~M Will.
\newblock Special relativity: a centenary perspective.
\newblock In {\em Einstein, 1905--2005}, pages 33--58. Springer, 2005.
\bibitem{jacobson2006lorentz}
Ted Jacobson, Stefano Liberati, and David Mattingly.
\newblock Lorentz violation at high energy: concepts, phenomena, and astrophysical constraints.
\newblock {\em Annals of Physics}, 321(1):150--196, 2006.
\bibitem{wolf2006recent}
Peter Wolf, Sebastien Bize, Michael~E Tobar, Frederic Chapelet, Andre Clairon,
  Andr{\'e}~N Luiten, and Giorgio Santarelli.
\newblock Recent experimental tests of special relativity.
\newblock In {\em Special Relativity}, pages 451--478. Springer, 2006.
\bibitem{mattingly2008have}
David Mattingly.
\newblock Have we tested lorentz invariance enough?
\newblock {\em arXiv preprint arXiv:0802.1561}, 2008.
\bibitem{minkowski1909raum}
Hermann Minkowski.
\newblock Raum und zeit.
\newblock {\em Jahresbericht der deutschen Mathematiker-Vereinigung},
  18:75--88, 1909.
\bibitem{von1910einige}
WA~Von~Ignatowsky.
\newblock Einige allgemeine bemerkungen zum relativit{\"a}tsprinzip.
\newblock {\em Verh. Deutsch. Phys. Ges}, 12:788--796, 1910.
\bibitem{frank1911transformation}
Philipp Frank and Hermann Rothe.
\newblock {\"U}ber die transformation der raumzeitkoordinaten von ruhenden auf
  bewegte systeme.
\newblock {\em Annalen der Physik}, 339(5):825--855, 1911.
\bibitem{frank1912herleitung}
Philipp Frank and Hermann Rothe.
\newblock Zur herleitung der lorentztransformation.
\newblock {\em Physikalische Zeitschrift}, 13:750--753, 1912.
\bibitem{schwartz1962axiomatic}
HM~Schwartz.
\newblock Axiomatic deduction of the general Lorentz transformations.
\newblock {\em American Journal of Physics}, 30(10):697--707, 1962.
\bibitem{lee1975lorentz}
Anthony~R Lee and Tomas~M Kalotas.
\newblock Lorentz transformations from the first postulate.
\newblock {\em American Journal of Physics}, 43(5):434--437, 1975.
\bibitem{levy1976one}
Jean-Marc L{\'e}vy-Leblond.
\newblock One more derivation of the Lorentz transformation.
\newblock {\em American Journal of Physics}, 44(3):271--277, 1976.
\bibitem{schwartz1984deduction}
HM~Schwartz.
\newblock Deduction of the general Lorentz transformations from a set of necessary assumptions.
\newblock {\em American Journal of Physics}, 52(4):346--350, 1984.
\bibitem{mermin1984relativity}
N~David Mermin.
\newblock Relativity without light.
\newblock {\em American Journal of Physics}, 52(2):119--124, 1984.
\bibitem{schwartz1985simple}
HM~Schwartz.
\newblock A simple new approach to the deduction of the Lorentz transformations.
\newblock {\em American Journal of Physics}, 53(10):1007--1008, 1985.
\bibitem{singh1986lorentz}
Sardar Singh.
\newblock Lorentz transformations in Mermin's relativity without light.
\newblock {\em American Journal of Physics}, 54(2):183--184, 1986.
\bibitem{sen1994galileo}
Achin Sen.
\newblock How galileo could have derived the special theory of relativity.
\newblock {\em American Journal of Physics}, 62(2):157--162, 1994.
\bibitem{Field_2001}
J.~H. Field.
\newblock Space-time exchange invariance: Special relativity as a symmetry principle.
\newblock {\em American Journal of Physics}, 69(5):569–575, May 2001.
\bibitem{pal2003nothing}
Palash~B Pal.
\newblock Nothing but relativity.
\newblock {\em European journal of physics}, 24(3):315, 2003.
\bibitem{coleman2003dual}
Brian Coleman.
\newblock A dual first-postulate basis for special relativity.
\newblock {\em European journal of physics}, 24(3):301, 2003.
\bibitem{field2004new}
John~Henry Field.
\newblock A new kinematical derivation of the Lorentz transformation and the particle description of light.
\newblock {\em arXiv preprint physics/0410262}, 2004.
\bibitem{dunstan2008derivation}
DJ~Dunstan.
\newblock Derivation of special relativity from Maxwell and Newton.
\newblock {\em Philosophical Transactions of the Royal Society A: Mathematical,
  Physical and Engineering Sciences}, 366(1871):1861--1865, 2008.
\bibitem{pelissetto2015getting}
Andrea Pelissetto and Massimo Testa.
\newblock Getting the Lorentz transformations without requiring an invariant speed.
\newblock {\em American Journal of Physics}, 83(4):338--340, 2015.
\bibitem{mathews2020seven}
WN~Mathews~Jr.
\newblock Seven formulations of the kinematics of special relativity.
\newblock {\em American Journal of Physics}, 88(4):269--278, 2020.
\bibitem{heaviside2003electromagnetic}
Oliver Heaviside.
\newblock {\em Electromagnetic theory}, volume 237.
\newblock American Mathematical Soc., 2003.
\bibitem{heaviside2011electrical}
Oliver Heaviside.
\newblock {\em Electrical papers}, volume~2.
\newblock Cambridge University Press, 2011.
\bibitem{heaviside1892xi}
Oliver Heaviside.
\newblock Xi. on the forces, stresses, and fluxes of energy in the electromagnetic field.
\newblock {\em Philosophical Transactions of the Royal Society of London.(A.)},
  (183):423--480, 1892.
\bibitem{larmorwork}
Joseph Larmor.
\newblock {\em Collected Papers}, 1928.
\bibitem{polchinski1996string}
Joseph Polchinski.
\newblock String duality.
\newblock {\em Reviews of Modern Physics}, 68(4):1245, 1996.
\bibitem{horowitz2009gauge}
Gary~T Horowitz and Joseph Polchinski.
\newblock Gauge/gravity duality.
\newblock {\em Approaches to quantum gravity}, pages 169--186, 2009.
\bibitem{padmanabhan1997duality}
Thanu Padmanabhan.
\newblock Duality and zero-point length of spacetime.
\newblock {\em Physical review letters}, 78(10):1854, 1997.
\bibitem{PhysRev.15.269}
Arthur~C. Lunn.
\newblock A principle of duality in thermodynamics.
\newblock {\em Phys. Rev.}, 15:269--276, Apr 1920.
\bibitem{landau1976mechanics}
Lev~Davidovich Landau and Evgenii~Mikhailovich Lifshitz.
\newblock {\em Mechanics: Volume 1}, volume~1.
\newblock Butterworth-Heinemann, 1976.
\bibitem{resnick1971introduction}
Robert Resnick.
\newblock Introduction to special relativity.
\newblock 1971.
\bibitem{eisenberg1967necessity}
Leonard~J Eisenberg.
\newblock Necessity of the linearity of relativistic transformations between inertial systems.
\newblock {\em American Journal of Physics}, 35(7):649--649, 1967.
\bibitem{weinberg1972gravitation}
Steven Weinberg.
\newblock Gravitation and cosmology: principles and applications of the general theory of relativity.
\newblock 1972.
\bibitem{hamermesh2012group}
Morton Hamermesh.
\newblock {\em Group theory and its application to physical problems}.
\newblock Courier Corporation, 2012.
\bibitem{moore2003informal}
Guy Moore.
\newblock Informal lectures on lattice gauge theory.
\newblock {\em lecture notes, McGill University. URL: \url{https://theorie.ikp.physik.tu-darmstadt.de/qcd/moore/latt_lectures.pdf}} 2003.
\end{thebibliography}


\end{document}